\documentclass[preprint,journal]{vgtc}       

\ifpdf
  \pdfoutput=1\relax                   
  \usepackage{graphicx}                
  \DeclareGraphicsExtensions{.pdf,.png,.jpg,.jpeg} 
\else
  \usepackage{graphicx}                
  \DeclareGraphicsExtensions{.eps}     
\fi%

\graphicspath{{figures/}{pictures/}{images/}{./}} 

\usepackage{microtype}                 
\PassOptionsToPackage{warn}{textcomp}  
\usepackage{textcomp}                  
\usepackage{mathptmx}                  
\usepackage{times}                     
\usepackage{cite}                      
\usepackage{tabu}                      
\usepackage{booktabs}                  

\onlineid{1217}

\vgtccategory{Research}
\vgtcpapertype{Application/Design Study}

\usepackage{subcaption}

\usepackage{xcolor}
\definecolor{light-gray}{gray}{0.95}
\definecolor{amaranth}{rgb}{0.9, 0.17, 0.31}
\usepackage{marginnote}
\usepackage[
  backgroundcolor=white,
  bordercolor=light-gray,
  linecolor=light-gray,
  textwidth=4em,
  textsize=footnotesize
]{todonotes}
\usepackage{xstring}
\newcommand{\note}[3]
{
\definecolor{author}{RGB}{103,0,31}
\IfSubStr{#1}{CF}{\definecolor{author}{RGB}{178,24,43}}{}
\IfSubStr{#1}{JW}{\definecolor{author}{RGB}{214,96,77}}{}
\IfSubStr{#1}{DK}{\definecolor{author}{RGB}{244,165,130}}{}
\IfSubStr{#1}{MW}{\definecolor{author}{RGB}{253,219,199}}{}
\IfSubStr{#1}{SK}{\definecolor{author}{RGB}{67,147,195}}{}
\IfSubStr{#1}{SC}{\definecolor{author}{RGB}{33,102,172}}{}
\IfSubStr{#1}{WW}{\definecolor{author}{RGB}{255,140,0}}{}
\ifx\submission\undefined
\textcolor{author}{%
#3%
\todo[
]{#1: #2}%
}%
\else
#3%
\fi
}%

\usepackage{enumitem}
\usepackage{soul}

\newcommand{\add}[1]{#1}
\newcommand{\rem}[1]{}
\newcommand{\cam}[1]{#1}

\usepackage{pgfplotstable}
\usepackage{booktabs,colortbl} 
\usepackage{tabularx}
\usepackage{hyperref}

\usepackage{caption}

\makeatletter\def\RemoveSpaces#1{\zap@space#1 \@empty}\makeatother

\newcommand{\CDataUpdate}{C1\add{--\textit{Changes}}}
\newcommand{\CPDataUpdate}{C1\add{~(\textit{Changes})}}
\newcommand{\ChallengeDataUpdate}{C1.~Adapting to Data Changes}

\newcommand{\CEdgeCases}{C2\add{--\textit{Edge Cases}}}

\newcommand{\ChallengeEdgeCases}{C2.~Anticipating Edge Cases}

\newcommand{\CTechnicalChallenges}{C3\add{--\textit{Constraints}}}

\newcommand{\ChallengeTechnicalChallenges}{C3.~Understanding Technical Constraints}

\newcommand{\CDataInteraction}{C4\add{--\textit{Interactions}}}

\newcommand{\ChallengeDataInteraction}{C4.~Articulating Data-Dependent Interactions}

\newcommand{\CDataMapping}{C5\add{--\textit{Data Mappings}}}
\newcommand{\CPDataMapping}{C5\add{~(\textit{Data Mappings})}}
\newcommand{\ChallengeDataMapping}{C5.~Communicating Data Mappings}

\newcommand{\CIterationsDiff}{C6\add{--\textit{Integrity}}}
\newcommand{\CPIterationsDiff}{C6\add{~(\textit{Integrity})}}
\newcommand{\ChallengeIterationsDiff}{C6.~Preserving Data Mapping Integrity across Iterations}

\newcommand{\OpportunityDataMapping}{Representing Data Mappings}
\newcommand{\OpportunityDataSketching}{Sketching Data-driven Visualizations}
\newcommand{\OpportunityDataInteraction}{Prototyping Data-Driven Interaction}
\newcommand{\OpportunityIterationsDiff}{Revealing Differences between Versions}
\newcommand{\OpportunityDataOutliers}{Testing Designs against Outliers in Data}
\newcommand{\OpportunityDesignDoc}{Authoring Data-driven Design Documents}

\newcommand{\ChallengeSection}[1]{\subsubsection*{#1}}

\newcommand{\OpportunityColor}[1]{%
\definecolor{OpportunityColor}{RGB}{255,255,255}
\IfSubStr{\RemoveSpaces{#1}}{\RemoveSpaces{\OpportunityDataMapping}}{\definecolor{OpportunityColor}{RGB}{166,206,227}}{}%
\IfSubStr{\RemoveSpaces{#1}}{\RemoveSpaces{\OpportunityDataSketching}}{\definecolor{OpportunityColor}{RGB}{31,120,180}}{}%
\IfSubStr{\RemoveSpaces{#1}}{\RemoveSpaces{\OpportunityDataInteraction}}{\definecolor{OpportunityColor}{RGB}{178,223,138}}{}%
\IfSubStr{\RemoveSpaces{#1}}{\RemoveSpaces{\OpportunityIterationsDiff}}{\definecolor{OpportunityColor}{RGB}{51,160,44}}{}%
\IfSubStr{\RemoveSpaces{#1}}{\RemoveSpaces{\OpportunityDataOutliers}}{\definecolor{OpportunityColor}{RGB}{251,154,153}}{}%
\IfSubStr{\RemoveSpaces{#1}}{\RemoveSpaces{\OpportunityDesignDoc}}{\definecolor{OpportunityColor}{RGB}{227,26,28}}{}%
}%

\usepackage{tikz}

\hypersetup{bookmarksdepth=3} 

\hypersetup{hidelinks=true} 

\title{Data Changes Everything: \\Challenges and Opportunities in Data Visualization Design Handoff}

\author{%
Jagoda Walny,
Christian Frisson, %
Mieka West, %
Doris Kosminsky, \\
S{\o}ren Knudsen, %
Sheelagh Carpendale, %
Wesley Willett%
}
\authorfooter{
\item
 All authors are with the University of Calgary, Calgary, Alberta, Canada.
 \item
 Christian Frisson's secondary affiliation is with McGill University, Montreal, Quebec, Canada. 
 E-mail: christian@frisson.re.
\item
 Doris Kosminsky's primary affiliation is with Federal University of Rio de Janeiro, Rio de Janeiro, Brazil. 
 E-mail: doriskos@eba.ufrj.br.
\item
 S{\o}ren Knudsen's secondary affiliation is with University of Copenhagen, Copenhagen, Denmark. 
 E-mail: sknudsen@ucalgary.ca.
\item
 Sheelagh Carpendale's primary affiliation is with Simon Fraser University, Vancouver, British Columbia, Canada. 
 E-mail: sheelagh@sfu.ca.
\item
 Remaining author e-mails: \{jkwalny, mieka.west, wesley.willett\}@ucalgary.ca.%
}
\shortauthortitle{Walny \MakeLowercase{\textit{et al.}}: Data Changes Everything}

\abstract{ 
Complex data visualization design projects often entail collaboration between people with different visualization-related skills. For example, many teams include both designers who create new visualization designs and developers who implement the resulting visualization software.
We identify gaps between data characterization tools, visualization design tools, and development platforms that pose challenges for designer-developer teams working to create new data visualizations.
While it is common for commercial interaction design tools to support collaboration between designers and developers, creating data visualizations poses several unique challenges that are not supported by current tools. 
In particular, visualization designers must characterize and build an understanding of the underlying data, then specify layouts, data encodings, and other data-driven parameters that will be robust across many different data values. In larger teams, designers must also clearly communicate these mappings and their dependencies to developers, clients, and other collaborators. 
We report observations and reflections from five large multidisciplinary visualization design projects and highlight six data-specific visualization challenges for design specification and handoff. These challenges include \textit{adapting to changing data}, \textit{anticipating edge cases in data}, \textit{understanding technical challenges}, \textit{articulating data-dependent interactions}, \textit{communicating data mappings}, and \textit{preserving the integrity of data mappings across iterations}. Based on these observations, we identify opportunities for future tools for prototyping, testing, and communicating data-driven designs, which might contribute to more successful and collaborative data visualization design.
} 


\CCScatlist{ 
 \CCScat{H.5.2}{User Interfaces};%
}

\begin{document}


\firstsection{Introduction}

\maketitle

Creating \cam{custom} visualizations is a challenging, multifaceted problem that requires a combination of skills and tools for data analysis, design, and development. Designers and developers must gain an understanding of the dataset and its characteristics through data exploration, then design data mappings, aesthetics, and interactions based on it~\cite{bigelow_reflections_2014}. These designs also need to be realized and deployed, typically by writing software. Sometimes it is possible for one person to perform all of these activities given enough time and resources. However, for more complex visualization projects with limited timelines, it is more feasible to distribute these activities amongst people in specialized roles.

This distribution of roles creates the challenge of \textit{handoff}, the codifying and exchange of information between people working on different roles in a project, and the related challenge of communicating domain knowledge across roles. This problem is already well-known in general software design, where interaction designers are often distinct from software developers~\cite{maudet_design_2017}. Over the past two decades, a wide range of specialized tools has emerged to help interaction designers outline and prototype interfaces in ways that reduce the friction between graphical designs and code. Commercial tools like Adobe XD~\cite{adobexd}, InVision~\cite{invision}, and Sketch~\cite{sketch} support expressive and precise visual design, interactive prototyping of animations, transitions, interactions\cam{, and streamlined exporting of specifications and assets for collaborators.}

\begin{figure}[b!]
  \centering
  \includegraphics[width=\columnwidth]{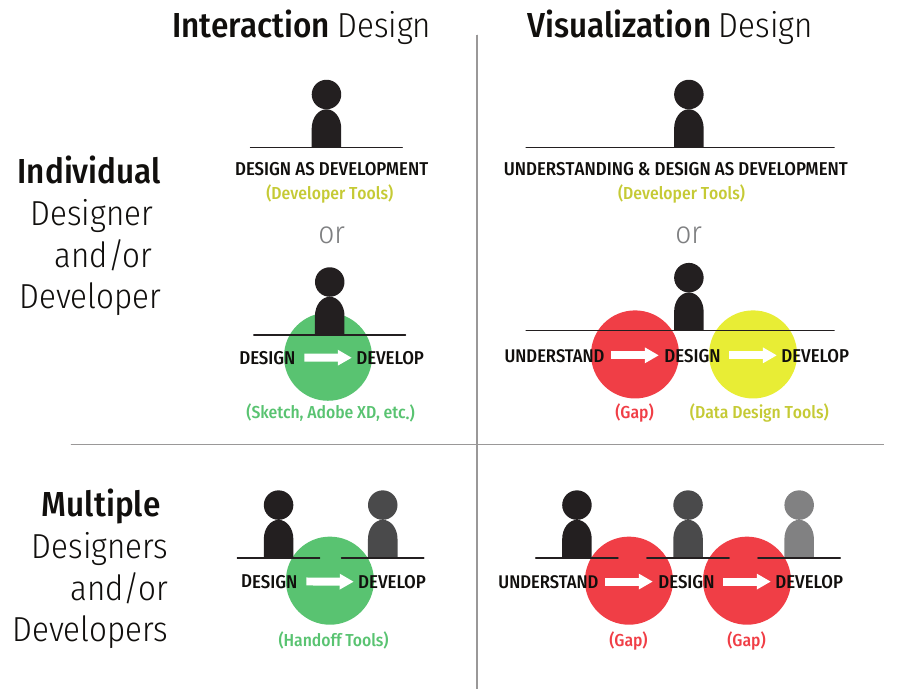}
  \caption{Contemporary interaction design tools increasingly enable smooth transitions and collaboration between design and development (top-left) and handoffs between designers and developers (bottom-left). In our experiences across projects, these transitions remain challenging for visualization designers (top-right) and teams (bottom-right). 
  }
  \label{fig:developersPlusDesign}
\end{figure}

\begin{figure*}[th!]
  \centering
  \includegraphics[width=\textwidth]{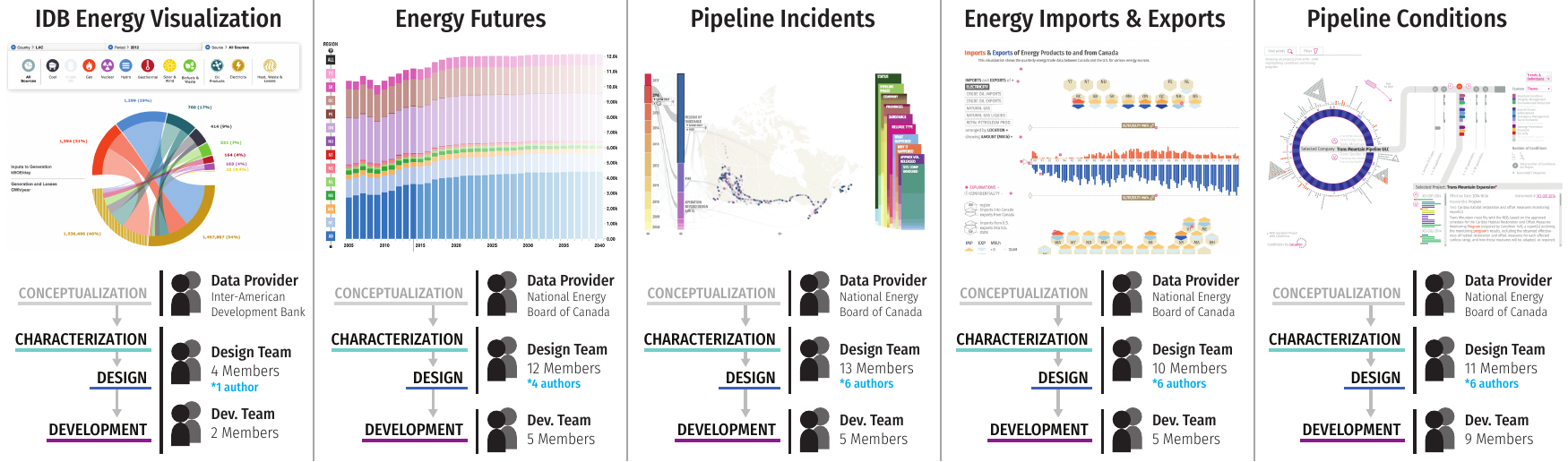}
  \caption{The five visualization design and development projects in which we ground our observations.}
  \vspace{-.2cm}
  \label{fig:projects}
\end{figure*}

Unfortunately for visualization designers, these tools lack robust support for data-driven designs. 
In practice, many programming-literate visualization designers still work largely in code, exploring datasets through iterative prototyping using libraries like D3~\cite{Bostock2011} or notebook environments like Observable~\cite{observablehq}. We \add{call} this  \emph{design-as-development}. However, \add{using these low-level tools} requires considerable technical skill and can increase the time and effort needed to articulate, refine, and polish visualization designs.
This scales poorly for large visualization projects, which may involve not just developers but also interaction designers, data experts, and clients, each with their own tool sets and institutional processes. In these collaborative settings, differing design objectives and gaps in the tools used to characterize data, design visualizations, and develop applications can exacerbate the issues caused by a lack of data-driven support. %
While unexpected issues can be handled with some agility by small, flexible teams, \add{these issues are} easily amplified in larger teams.

We draw on our experience as part of the design team on five complex data visualization projects (Figure~\ref{fig:projects}) intended for wide-scale public release. 
Across each of these projects, we observed and participated in interactions between teams of designers and developers working together and separately to characterize data, create initial designs, and translate those designs into production-ready applications.
\cam{Based on these experiences, we} highlight challenges and opportunities specific to designer-developer collaboration in data visualization design projects.

While others before us have discussed practical visualization design projects~\cite{kirby2013visualization, VandeMoere:2011Role} and design studies~\cite{Sedlmair:2012:DSM}, our focus is different. We focus on the practical work and coordination that goes into building visualizations. Although our projects did employ several researchers, they also employed \cam{design and development} practitioners. \add{Our projects also involved close coordination with project coordinators and data experts\cam{\footnote{\cam{Data experts had technical and/or domain-related data expertise.}}}
on the data provider's side.} This \add{complex structure and} the physical and temporal separation of different teams \add{heightened the visibility of several}  practical challenges still faced during the \add{visualization} design process. Our work articulates issues that are of a very practical nature and that we expect are frequently experienced by others \cam{engaged in} practical visualization work. We think our contributions add value to this practitioner-oriented research space, especially in light of the visualization research community's recent focus on practitioners as a crucial source of \textit{``energy, ideas, and application problems''}~\cite{restructuring}.

\section{Roles in Visualization Design Processes}
\label{sec:DataVisDesHandoff}

Visualization design and development requires several unique sets of skills including experience in
human-centered design, perception, evaluation, statistics, and graphics programming~\cite{kirby2013visualization}. 
This conventional wisdom
is exemplified by the (somewhat mythical) notion of ``full-stack''
visualization designer-developers capable of conducting the full range
of ``data wrangling, dynamic graphics, and
derring-do''~\cite{gray2012data}. 
However, real-world visualization design projects
(especially large ones) often include a variety of team members with
diverse and overlapping subsets of these skills. As projects grow, these
teams can become segmented, with responsibility for design and
development delegated to individuals or teams whose skill sets and
preferred tools can be increasingly disjoint. In particular,
institutional and disciplinary divides can result in the partitioning of
early-stage \emph{design} tasks --- such as data \cam{profiling}, ideation, creating mockups, and
prototyping --- and \emph{development} tasks like implementation,
testing, deployment, and maintenance.

\cam{In practice}, disciplinary divides between designers and developers \add{can be} stark~\cite{maudet_design_2017}.  The interaction design literature has examined the divide between designers and developers from a number of angles, including: how designers and developers align their work in collaborations~\cite{brown_joint_2012}, how they work remotely~\cite{Yiu_2014}, and how design tools can function as boundary objects that mitigate designers' lack of ``material'' experience with software~\cite{ozenc_how_2010}. Recent work by Maudet et al.~\cite{maudet_design_2017} has drawn attention to \emph{design breakdowns} \add{in design handoff}, in which potential disconnects between designers and developers are highlighted by difficulties in implementing the final design. \add{Leiva et al.~\cite{leiva2019enact} expand on this concept, identifying several specific types of breakdowns --- including \textit{omitting critical details}, \textit{ignoring edge cases}, and \textit{disregarding technical limitations} --- that routinely contribute to difficulties in projects involving handoffs between designers and developers.}

\begin{figure*}[t!]
  \centering
  \includegraphics[width=\textwidth]{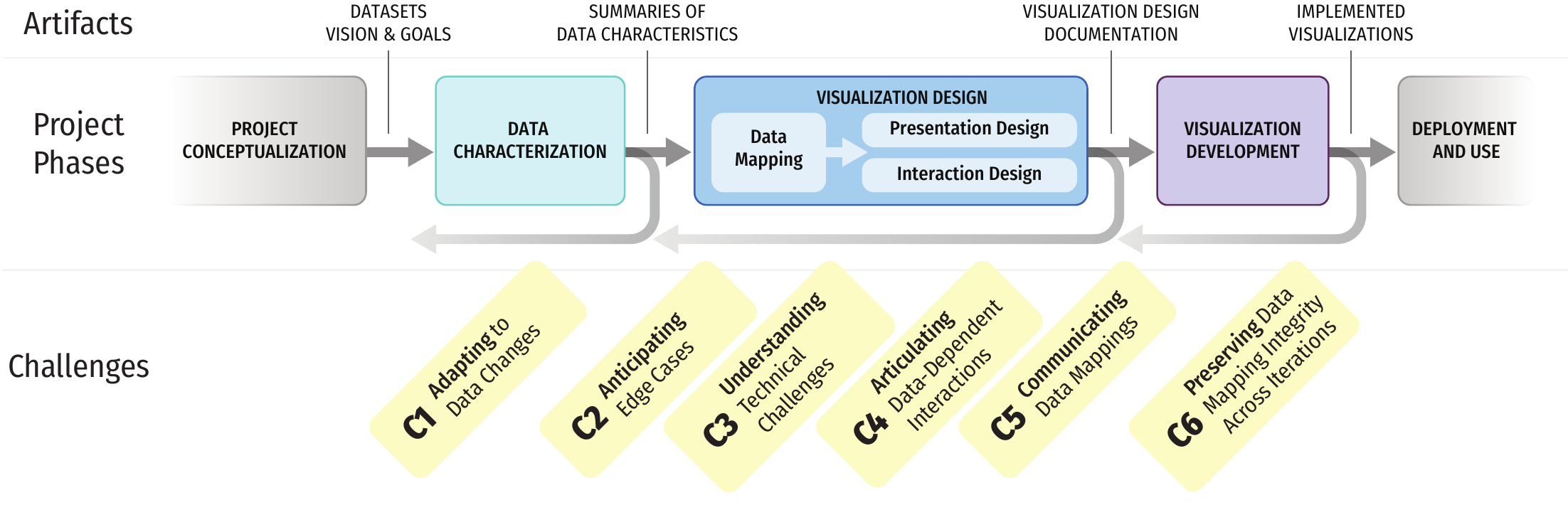}
  \caption{Stages of a data visualization development process and the dependencies between them. \cam{We focus on the \textit{data characterization}, \textit{design}, and \textit{development} phases, highlighting the artifacts that bridge these phases and the challenges (C1-C6) that emerge within them}.
  }
  \vspace{-0.2cm}
  \label{fig:dependency-chain}
\end{figure*}

However, research in data visualization design still often fails to acknowledge this division of design and development. 
\add{For example,} McKenna et al.'s Design Activity Framework~\cite{McKenna2014} combines design and implementation into a
single ``make'' step and assumes that the responsibility for both will
be tightly integrated. Other reflections on visualization design
practice also tend to share this assumption, drawing primarily on the
perspectives of visualization design researchers tasked with \emph{both
creating} \emph{and implementing} novel visualizations \cam{for multidisciplinary collaborators}~\cite{kirby2013visualization,Sedlmair:2012:DSM,VandeMoere:2011Role}.

The design and development of new visualizations, like that of other interactive systems, entails considerable iteration and involves transitions between multiple sets of tools as designs move from conception to implementation. 
While some amount of visualization design and development often happens via coded prototypes, many aspects of visualization design --- from early-stage concept generation and sketching through to late stage aesthetic refinement --- are often better-served by graphic design and interaction design tools that offer greater expressive flexibility, as evidenced by the work practices of data visualization designers interviewed by Bigelow et al.~\cite{bigelow_reflections_2014}.
However, unlike most other work in interaction design, the form of new visualizations depends intrinsically on the data that they will communicate. As a result, the process of visualization design is often a complex and iterative one anchored in multiple rounds of data examination, ideation, creation, and deployment~\cite{McKenna2014}.
These activities pose challenges for designers who may need to transition repeatedly between interactive tools that allow them to examine data and explore a diverse range of designs and more low-level data-driven development and coding. 

These issues are compounded as projects grow and responsibilities for design, development, and deployment are divided across multiple individuals or teams, each with different skill sets and priorities. 
In \add{these} situations, visualization design and development become an exercise in
co-creation~\cite{bigelow_reflections_2014}, complicated by dependencies between teams and differences in their competencies. Large diverse teams make it possible to create, deploy, and provide long-term support for complex visualizations. However, this division of labor reveals a variety of new design handoff and iteration challenges, which can be \cam{compounded} by the data-driven nature of visualization design.


\section{Overview of Visualization Design Projects}

Our reflections on handoff in visualization design and development are anchored in our own experiences as members of a design team on five large data visualization design projects (Figure~\ref{fig:projects}) conducted between 2012 and 2019. Each project \cam{was intended for wide-scale public release} and involved between six months and several years of data characterization, design, and development work.

For each project, the work was directed by an outside client who was also the data provider.  Our multi-member \textbf{design teams}, which included a rotating cast of designers, visualization researchers, post-docs, graduate students, and interns,  were responsible for the majority of the data characterization and visualization design. In all projects, at least one (and typically more) of the authors participated in the process directly as members of the design team. A separate \textbf{development team} was tasked with creating, deploying, and providing initial maintenance for the final web-based applications.

\subsection{Projects}
\noindent\textbf{Energy Visualization for the Inter-American Development Bank (IDB).}
The earliest of the projects, conducted between 2012--16 with the Inter-American Development Bank produced a suite of visualizations showcasing energy source generation, import and export, transmission, and consumption for countries in the Americas, as well as other benchmark countries.  The resulting visualizations were hosted publicly from 2013--18, but are no longer accessible as of 2019. In this project, the design and development teams were more closely integrated than in the other projects, with both playing a substantial role in data characterization, design, and development.

\noindent\textbf{Energy Futures.}
This project, conducted over %
4 months in 2016, led to the development of four visualizations based on forecasts of Canadian energy production and consumption~\cite{Blascheck_2018}. A second %
7-month iteration~\cite{Knudsen_2018} of the project in 2017 added a fifth visualization showcasing changes in projected energy demand across the Canadian provinces and territories. 
The visualizations are publicly available at \href{https://apps2.neb-one.gc.ca/dvs}{https://apps2.neb-one.gc.ca/dvs}.

\noindent\textbf{Pipeline Incidents.}
Developed during 2017, this 8-month project produced an interactive visualization system that supported visual exploration of incidents that occurred on or around federally-regulated pipelines. 
The visualization is publicly available at \href{https://apps2.neb-one.gc.ca/pipeline-incidents}{https://apps2.neb-one.gc.ca/pipeline-incidents}. 

\noindent\textbf{Energy Imports \& Exports.}
Another similarly-scoped project conducted over
16 months in 2017--18 involved creating a set of five visualizations showing historical imports and exports of various energy products from Canada. The visualizations are publicly available at 
\href{https://apps2.neb-one.gc.ca/imports-exports}{https://apps2.neb-one.gc.ca/imports-exports}.

\noindent\textbf{Pipeline Conditions.}
The most recent project, conducted \add{over 18 months} during 2018--19, focuses on visualizing the conditions placed by government regulators \add{on} the construction of new pipelines. 
\add{At the time of publication,} this project was near completion, but not yet publicly accessible.  

\subsection{Design Team Roles}
\add{The members of the design team needed to fulfill a variety of design-related roles. 
The project needed team members who could \textbf{characterize data} \cam{by wrangling data}, exploring data in existing visualization tools, spreadsheets, or code, processing data (including text mining), and understanding specific data types (for example, linguistic analysis of text data). 
All team members needed to \textbf{create and understand data mappings} from data to visual representation, which included varying degrees of ideation, basic perceptual understanding, and applying knowledge of visual variables. 
The project also required people with \textbf{visual design} skills who could design graphics, page layout, and typography while keeping accessibility in mind. The team included people with \cam{expertise} in \textbf{interaction design}, including prototyping and animation. Likewise, some team members \textbf{developed visualization prototypes} to verify and demonstrate designs and \cam{\textbf{engineered}} the technically complex portions of design documents. 
All team members needed to \textbf{collaborate} and \textbf{communicate} with the data provider and development team. 
} 

\add{During the Inter-American Development Bank project, the design team consisted of one primary visualization design researcher and three Visual Arts students (one undergraduate and two graduate students). The development team consisted of one primary computer science researcher and one doctoral student. These teams worked closely together in an iterative fashion and were located on the same campus.}

\add{In the remaining projects, the design and development teams were separate. The design team consisted of two primary investigators (visualization researchers), one project coordinator, one design researcher, \cam{1--2} 
postdoctoral visualization researchers, 2--3 undergraduate or recently graduated computer science students, 0--1 information design undergraduate students, and 1--4 full-time employees with roles in design, development, and specialized data understanding. The development team consisted of anywhere between 5 and 9 members of a professional software development firm located in the same city.}
\add{In all \cam{projects}, the data providers were from separate institutions and were physically separated from the design and development teams. }

\subsection{Analysis and Synthesis Process}
\add{We identified the data-related challenges described in this paper via an ongoing process of reflection~\cite{meyer_reflection_2018} through which we worked to refine our design team's work and communication practices.} 
During each project, we kept records of artifacts produced for meetings, data explorations, ideation sketches, planning timelines, and design documents. We also regularly reflected on communication and design challenges within our own team. 
\add{Throughout, we documented and scrutinized the process using approaches drawn from diary-based~\cite{Czerwinski:2004:DST:985692.985715,ohly2010diary,Rieman:1993:DSW:169059.169255,sonnentag2001work} and autobiographical studies~\cite{bullough2001guidelines,desjardins2018revealing,neustaedter2012autobiographical}. Individual members of the design team, as part of their personal practice, kept notes and images documenting their work. Later, as part of this autobiographical process, we carefully re-examined our diary-based records and used them to identify gaps and challenges.}

\add{\cam{We} also took steps to formalize our design communication processes. Within the design team, we leveraged our initial observations to create shared tools and processes for tracking the team's progress.} 
\add{During the implementation phase of each project, we also held face-to-face design review meetings with all stakeholders present.  Finally, after each of the three \cam{later} projects, we organized formal process discussions with the data provider and with members of both the design and development teams to help improve coordination for subsequent projects. We took detailed collaborative notes at all meetings.}

\add{Based on these reflections, we focused increasing energy across the remaining projects on improving design communication both within and across the teams. As part of this effort we documented meetings and design processes using  detailed personal records, team records, design handoff documents, and handoff document revisions. We teased apart the details of the challenges presented in this paper by drawing on these detailed records. In discussing a particular challenge we could rigorously re-examine the time-stamped process by which each design was created, \cam{handed off}, implemented, re-discussed, re-implemented, and ultimately released. 

}

\add{Throughout our reflection, we noticed that a number of the recurring handoff challenges were not merely interaction design issues (like those documented by Leiva et al.~\cite{leiva2019enact}), but were instead rooted in the deep dependence of the designs on data.} From these reflections, we have synthesized the most prominent unresolved \add{data-related} challenges and illustrated them using real examples from our projects. Where possible, the initial reflection was \cam{drafted} by the team member who most closely experienced the example issue.

\section{Visualization Design \& Development}
\label{sec:VisDesignDevelopment}

One outcome of our reflection on the processes and communication in our design projects is a formal model of the major phases of our design projects (Figure~\ref{fig:dependency-chain}). This model was born out of a need for a vocabulary to use when coordinating with multiple parties, and serves as a useful anchor for the design and communication challenges we discuss in the remainder of this paper. 
In a designer-as-developer scenario, a single person or small team might carry out all of these phases with less need for a formal process. In contrast, in our scenarios \cam{---} which \cam{largely} involved multiple teams who did not share a daily working space \cam{---} this model emphasizes the distribution of roles \cam{ and responsibilities}. In addition, it highlights the kinds of artifacts that can often facilitate communication across phases. 

\noindent\textbf{Project Conceptualization.} This phase occurred on the client side, prior to the direct involvement of the design team. The client provided the design team with the \textbf{\emph{vision and goals}} for the project as well as a \textbf{\emph{dataset}}. The handoff of these artifacts ranged from simple emailed delivery to more involved full-day workshops between client-side data experts and the design and development teams. 

\noindent\textbf{Data Characterization.} In this phase, the design team explored and characterized the data, prioritizing analyses motivated by the project vision. This included understanding data types, amounts, and extrema\cam{,  as well as} the relation\cam{ship between the data and} the project goals. In some cases, the design team recommended a more focused dataset for the visualization, \cam{a process which sometimes entailed} several iterations with the client-side data team. \add{This phase \cam{typically involved some} data wrangling~\cite{kandel_2012_interview}, but was more akin to exploratory data analysis or domain problem characterization~\cite{munzner_nested_2009}.} 
We used a number of tools to support this phase, including spreadsheets, hand-coded scripts, \cam{and} visualization exploration software such as Tableau~\cite{Tableau}, \cam{as well as} hand-sketching for preliminary ideation. This culminated in the creation of a \textbf{\emph{summary of data characteristics}}, which \cam{we typically delivered} to the client-side data experts \cam{via an in-person presentation}. The \cam{insights and shared understanding gained} throughout this phase served as a foundation \cam{for future design work}. 

\noindent\textbf{Visualization Design.} The design phase \add{encompassed the abstraction and encoding/interaction design phases of the nested model~\cite{munzner_nested_2009} together with partial algorithm design and extensive visual presentation design.}
This was a two-stage process\cam{. F}irst, we developed a \cam{visualization concept and received initial approval from} the client. Then, we refined and polished the final design and documented it \cam{via a} \textbf{\emph{visualization design documentation}}  shared with both clients and developers. We developed a \emph{data mapping} on the basis of the data characterization\cam{. In this process, we relied} heavily on hand-sketching and manual illustration in tools like Adobe Illustrator~\cite{illustrator}. In some cases, we  used chart generation tools and utilities (\cam{including} RAWGraphs~\cite{mauri_rawgraphs:_2017} and Color Brewer~\cite{colorbrewer_2003}) and hand-coded prototypes using libraries such as D3~\cite{Bostock2011}. We developed the \emph{presentation design} --- the overall size and layout of the visualization --- primarily using Adobe Illustrator~\cite{illustrator}. Furthermore, we \cam{developed} the \emph{interaction \cam{design}} using various tools including paper prototyping~\cite{Rettig_1994_prototyping}, textual and sketched descriptions, and general-purpose interaction prototyping tools (\cam{such as} Axure~\cite{axure} and Atomic~\cite{atomic}). The final development document was initially in PDF form but later evolved to be a web-based document. The most recent design document was based on Idyll~\cite{conlen_idyll_2018}, which allowed us to combine mockups, coded prototypes, and explanatory text in a single document.

\noindent\textbf{Visualization Development.} This phase was led by the software development team. As the design team, our role was mainly reactionary\cam{:} we responded to questions about the design, suggested redesigns when issues arose, and verified that the implementation was functioning as intended. Most of the discussions \cam{in} this phase were grounded in the design documentation \cam{and in} increasingly polished iterations of the \textbf{\emph{implemented visualizations}}.

\noindent\textbf{Deployment and Use.} As the visualization was deployed for public use, the software development team was tasked with its maintenance, including implementing quarterly data updates. The design team was involved if a data update contained unexpected values that were not supported by the existing design.

\section{Challenges when Designing with Data}
\label{sec:challenges}

Based on our reflection and observation, we describe six gaps in the data visualization design process.

\ChallengeSection{\ChallengeDataUpdate}

\emph{Data updates can have cascading effects on the data characterization, visualization design, and development phases because all aspects of visualization development depend upon the data. The impact of such effects may not be clear to data providers.}

In our experience, data is rarely available in its full and final form before the visualization development process begins, often necessitating data updates later in the process --- sometimes even post-deployment. Data changes are particularly impactful if they change the data characterization or the data used to generate views in the visualization design. Even when a data change is seemingly innocuous and does not change the \cam{overall} data mapping, it may affect the implementation stage, particularly where server-side mechanisms for loading, aggregating, or preparing data have already been established. For example\cam{,} changing a column name or unit symbol may break existing data parsers.

Data updates are not necessarily undesirable. They might provide corrections or additional data\cam{, and} they might reflect a positive evolution \cam{in} how the data provider releases data for the visualization. \cam{This} evolution might itself be prompted by witnessing the interim results of the visualization design process. As such, the challenge is not to avoid data updates altogether but to be able to cope with them efficiently.

\begin{figure}[tb]
  \centering
  \includegraphics[width=\columnwidth]{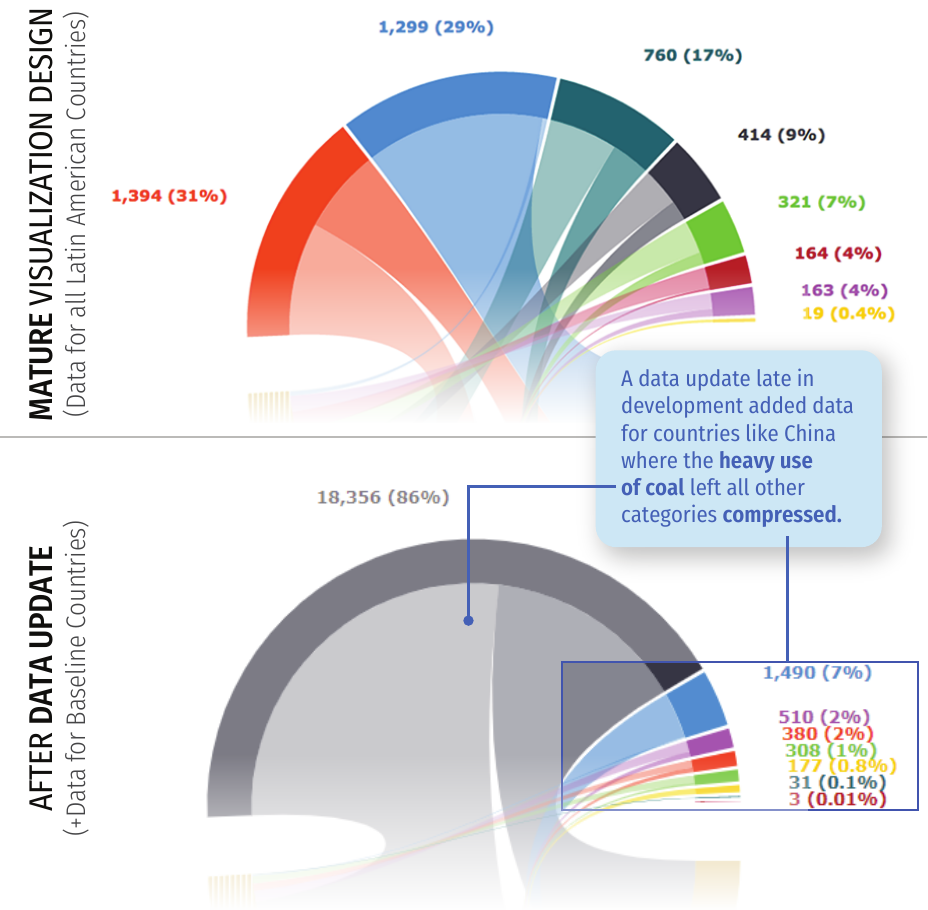}
  \caption{\add{Adding data from additional countries late in the design of this visualization resulted in a single energy source (coal) dominating the view which made other energy sources difficult to compare.}\rem{Impact of a major data update on a mature visualization design.}}
  \label{fig:datachange-iadb}
\end{figure}

For example, late in the process of designing visualizations for the \textit{Inter-American Development Bank} project, the design and development teams had created a mature, late-stage visualization design of energy generation data from Latin American countries (\autoref{fig:datachange-iadb}-top). Up to that point, all design decisions had been made on the basis of the team's work with the initial data \cam{that was} provided by the client. This characterization led to a design that arranged data about different energy sources on a circle, showing the relationship \cam{of} energy inputs (top half of the circle) to energy generation and losses (bottom half).

At this stage, an update added data for benchmark countries like China, creating views (\autoref{fig:datachange-iadb}-bottom) in which a single energy source (in this case coal) visually overwhelmed the values from other sources. This dramatically altered the form and legibility of the visualization, crowding all of the original data onto a small slice at the side of the chart, and overlapping the arcs and labels.  Given the late stage of the design process, there were not enough resources available to iterate the design, and the resulting visualization was quite different from the original concept.

\begin{figure}[b]
  \centering
  \includegraphics[width=\columnwidth]{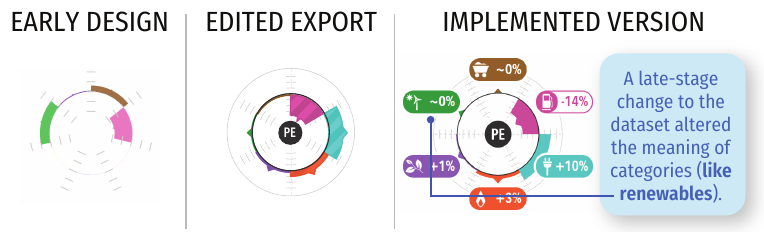}
  \caption{\add{Changes to the dataset for this \textit{Energy Futures} visualization altered category meanings, creating potentially misleading values.}}
  \label{fig:datachange-vis5}
\end{figure}

\add{Late-stage data updates can also cause subtle changes to how a visualization is perceived. For example, in the \textit{Energy Futures} project, a visualization of demand shares by energy source was initially designed using data based on the energy production stages. The design team characterized the data and moved on to the design phase, creating a D3 prototype (Figure~\ref{fig:datachange-vis5}). Our focus here was to support comparison between provinces that have order-of-magnitude differences in scale. Later, the data\cam{set} was changed to \cam{one} based on end-use energy demand. The characteristics of both datasets were quite similar, so the design work continued with attention shifting to other concerns. However, there was a key but subtle difference in the new dataset\cam{:} nearly all renewable energy was included within the Electricity category rather than in the Renewables category. The Renewables category in the original dataset was already quite small, so this change went unnoticed. While the resulting visualization still shows the data accurately, it \cam{can be easily misinterpreted because it does not clearly} communicate the fact that the Electricity category \cam{ also includes most renewables.}}

\ChallengeSection{\ChallengeEdgeCases}

\emph{It is difficult for designers to anticipate and test all possible combinations of interactive inputs that a visualization might receive. As a result, it can be hard to anticipate situations in which a chart design or data mapping may break.}

Common visualization interactions such as filtering or aggregation \cam{essentially}  change data mappings in real time. In design-as-development scenarios with live prototypes, these kinds of interactions can be tested relatively early in the development process and designs can be adjusted as needed. However, when design and development are separated, designers do not always have the tools or skillsets to fully test all possible combinations of inputs. As such, potential problems might only be uncovered during the development phase after many design decisions have been finalized. At this stage, any changes to the design can incur significant design and development work, limiting the possible ways in which the visualization design can be adapted to mitigate the problems. 

\begin{figure}
  \centering
  \includegraphics[width=\columnwidth]{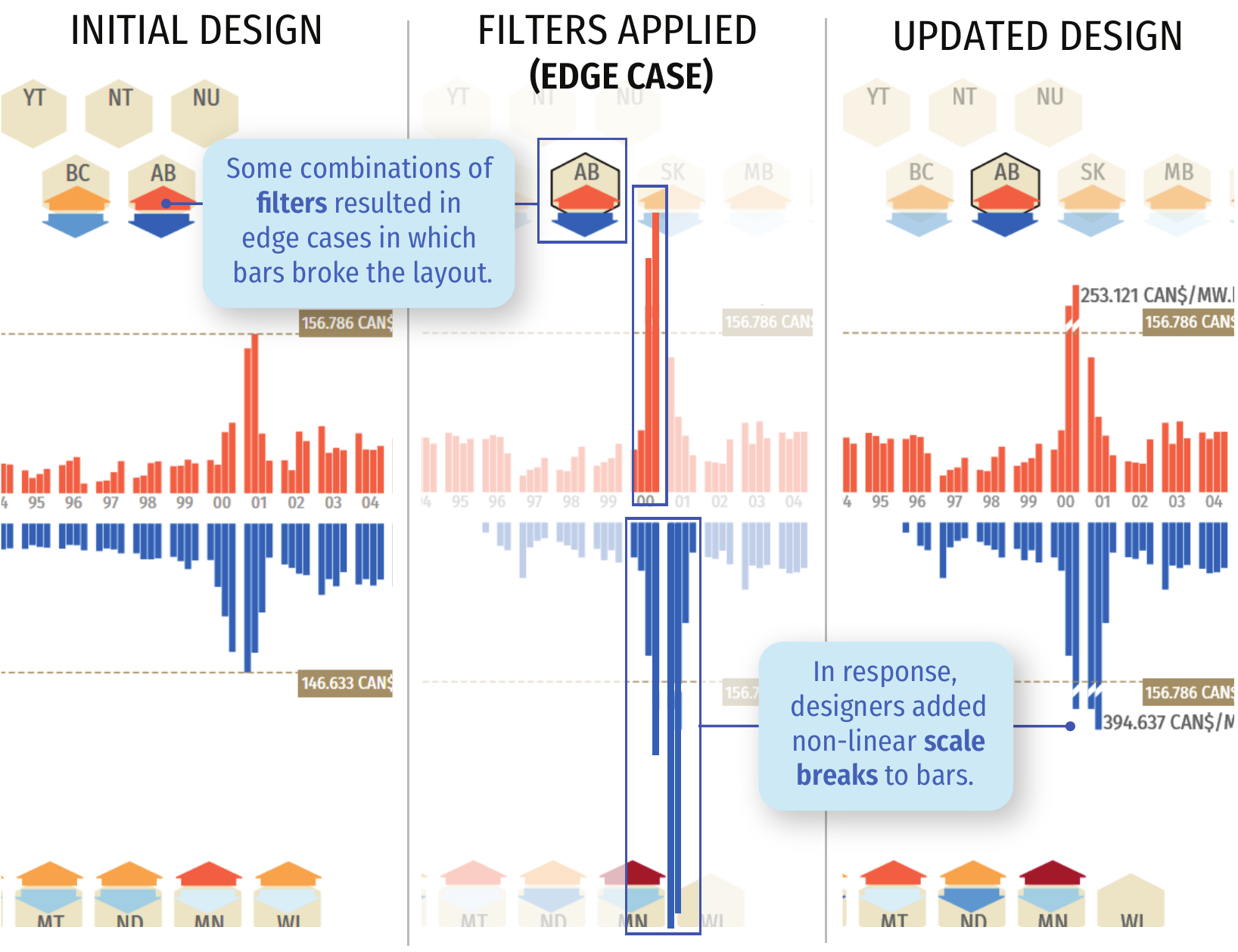}
  \caption{The initial design of the \textit{Energy Imports \& Exports} visualization~(left) responded poorly to particular combinations of filters (center) and ultimately required a revision (right).}
  \label{fig:edgecases}
\end{figure}

In the design of the \textit{Energy Imports \& Exports} visualizations, the design team developed a mirrored chart that could be filtered to show average quarterly electricity prices between any combination of US states and Canadian Provinces (\autoref{fig:edgecases}-left). This data mapping worked well for the vast majority of views, including the various combinations of test data that the design team used when creating their initial documents. However, once this design was implemented it became clear that filtering by particular combinations of states and provinces revealed outliers which had been masked in the initial samples (\autoref{fig:edgecases}-center).

The design solution was constrained by the fixed size and very limited space available for the bar chart, as well as by the need to maintain consistency with the data mapping in other parts of the visualization. Reconfiguring other parts of the visualization was also not feasible late in the development stage.  
Ultimately, the design team opted to use a compressed scale break for these outliers (Figure~\ref{fig:edgecases}-right). This solution makes it impossible to make direct visual comparisons between  large values above the scale break and reduces the visual impact of large values, but still communicates relative differences in scale within the available space and minimize\cam{s} changes in other parts of the visualization. \add{Several related scale issues also emerged late in the development of the \textit{Pipeline Conditions} visualization, necessitating new design work during implementation. In all cases, if the edge cases had been identified} earlier in the design process, the entire design might have been conceived differently.

\ChallengeSection{\ChallengeTechnicalChallenges}

\emph{Designers may not be aware of all of the technical constraints and challenges that can occur during the development phase. This \cam{can lead} to uncertainty about design feasibility and can \cam{also} trigger more dramatic revisions during development.}

It is not always clear what types of software and hardware limitations may pose challenges for a design. Visualization designers tend to focus considerable attention on the choice of visual mappings and on providing a useful and appealing interaction experience. Web and application developers, meanwhile are more likely to be tasked with delivering robust, efficient, standards-compliant implementations of a design. While collaborators often have an appreciation for the others' areas of focus, it can be difficult to be aware of all potential issues that the other group faces. From a design perspective, this can lead to uncertainty, particularly when proposing unique or unconventional designs. Across our projects, the designers often tried to mitigate this uncertainty by prototyping and testing novel pieces of the designs in code. However, we still frequently encountered technical hiccups during development. 

\add{For example, during the design of the \emph{Pipeline Conditions} visualization\cam{,} the design team created detailed working versions of several complex components, including an interactive keyword browser that leveraged a third-party physics engine. However, the development team worked within a different set of constraints that included cross-browser compatibility, future code maintainability, and a decision to use a different underlying web framework to implement the site. As such, they chose to re-implement the\cam{se} components from scratch. This resulted in controls that were superficially similar to the original designs but \cam{that} behaved quite differently. As a result, they required substantial additional refinement to achieve behaviour that was already present in the \cam{design team's} prototypes.
}

\begin{figure}[tb]
  \centering
  \includegraphics[width=\columnwidth]{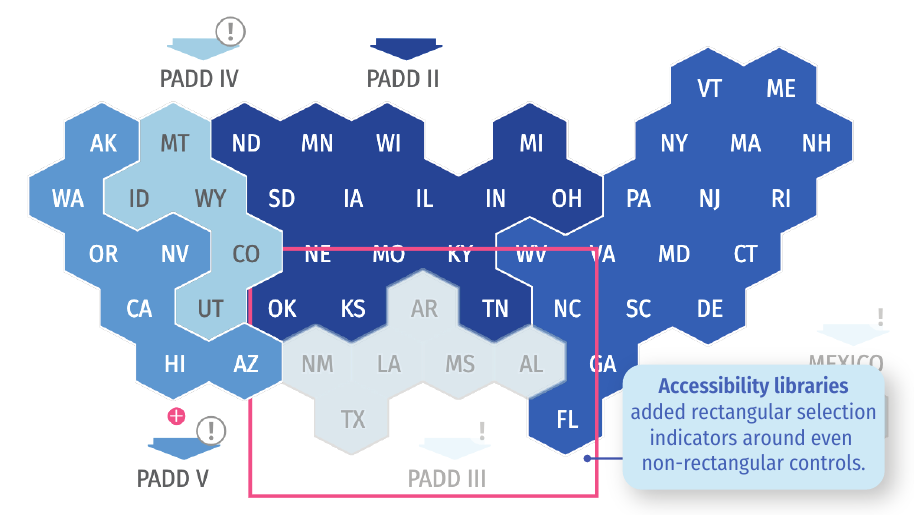}
  \caption{A portion of a visualization showing a heat map of Canadian energy exports to United States Petroleum Administration Defense Districts (PADDs). Due to confusion surrounding the limitations of an accessibility template, color mappings and the layouts of non-rectangular components in this visualization needed to be adapted to accommodate rectangular selection indicators.}
  \label{fig:technicalConstraints}
\end{figure}
Often, these issues arise not from limitations in the visualization libraries themselves, but in visualization-adjacent aspects of the implementation such as data loading, cross-browser compatibility, and accessibility requirements. 
During the development phase of the \textit{Energy Imports \& Exports} project, the design team was surprised to learn that the government-mandated accessibility template within which the design would sit also applied to individual components of the visualization. This meant that each selectable element of the visualization would be surrounded by a stroked rectangular box with a dominant color. This late discovery resulted in a mismatch between the visual aesthetic of the design, which was built around hexagonal tiles, and the bright halos produced by the accessibility template (\autoref{fig:technicalConstraints}). The use of the template also forced the design team to revisit the data mapping to ensure that the color used by the accessibility overlays did not conflict with the palette used to encode import and export data.


\ChallengeSection{\ChallengeDataInteraction}

\emph{When ideating and specifying new data-dependent interactions, designers often need to generate a variety of different data-accurate views showing the visualization in multiple states. This extra cost can make data-driven interactions challenging to develop and \cam{communicate} to other team members.}

The outcome of most interactions with a visualization \cam{depends} on the data itself. Typically, these data-dependent interactions generate new views of the visualization that may represent a different subset of the data, a different transformation of the data, or a different encoding of the data. For instance, filtering operations reduce the set of data being considered. This can result in a change of position or appearance of the remaining elements. Similarly, each brushing and linking \cam{interaction} requires a change to the encoding applied to a very specific set of marks spread across several views. When prototyping these interactions by hand or using graphic design tools, designers must manually manage and update large numbers of individual data elements. Modeling the transitions between such views is even more difficult, especially when complex animations are required. 

  \begin{figure}[tb]
    \centering
    \begin{subfigure}[b]{0.5\textwidth}
      \includegraphics[width=\columnwidth]{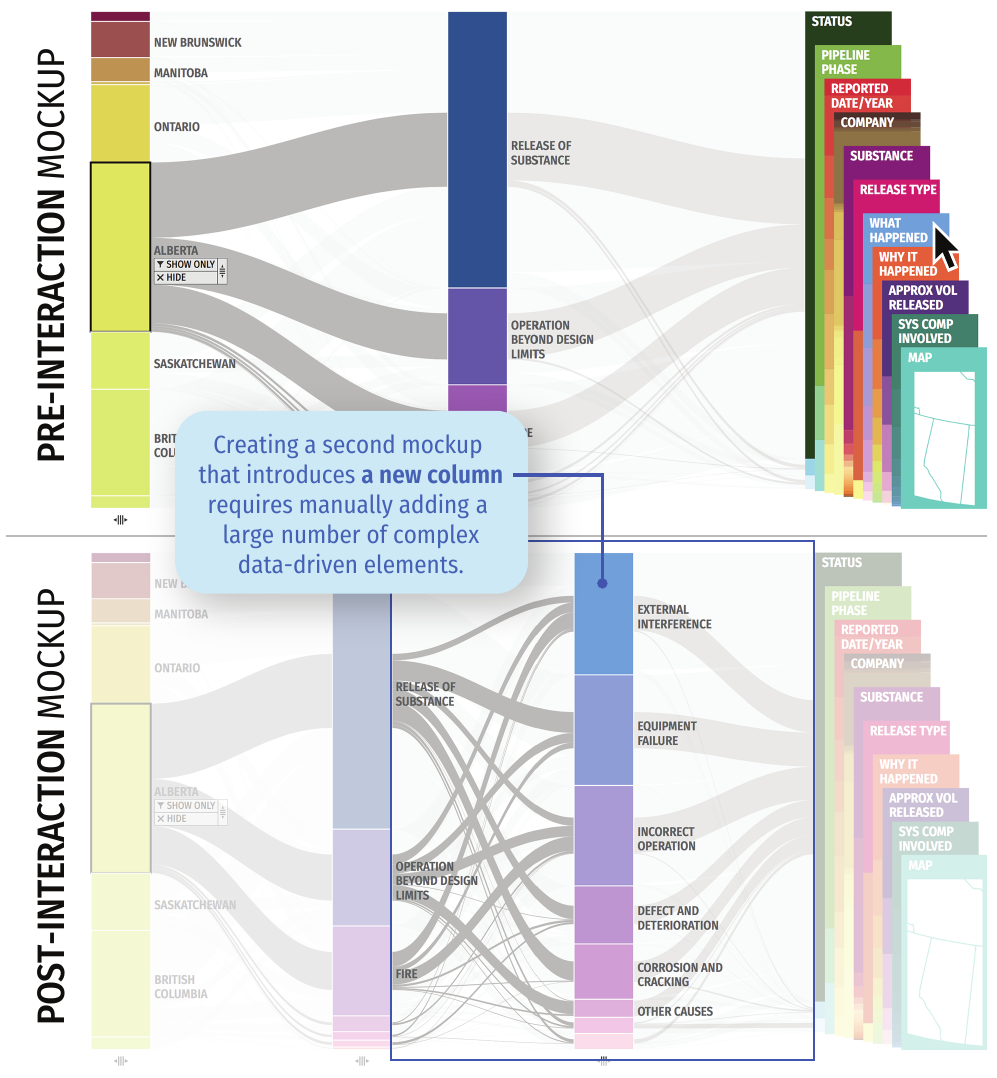}
      \label{fig:incidents-interaction}
    \end{subfigure}
    \caption{Two views of a parallel sets-style visualization of incidents on pipelines and pipeline facilities. \textbf{Top:} A single category is selected. Highlighted gray lines represent how the incidents in the selected category relate to incidents in the categories of the column to the right. \textbf{Bottom:} The design allows additional columns to be dragged into the visualization from the right. This single interaction introduces substantial complexity to the resulting view. Prototyping, testing, and communicating this resulting view in a data-accurate way is difficult using conventional graphic and interaction design tools.}
    \label{fig:incidents-interaction}
  \end{figure} 

Transitions like the one in \autoref{fig:incidents-interaction} are particularly challenging. In this example, drawn from our \textit{Pipeline Incidents} project, a simple interaction with the flow visualization can add an additional column, introducing dozens or even hundreds of new arcs. The complexity that this single interaction adds to the view is substantial. Multiple forking curves appear with varying positions and thicknesses, all of them related to the previous selection. 

Creating a data-accurate version of the resulting view is very difficult using conventional graphic design tools and requires manually computing the size, positions, and connectivity of large numbers of edges, then manually adding them to the mockup.  This expense is multiplied every time there is a change to the underlying design or an update to the data. Even trivial changes such as changing screen dimensions or color schemes can require manual revisions to these \cam{designs}. Specifying  transitions and animations between these views is also difficult, even when using interactive wireframing and animation tools, since they do not include data binding or animation support for such fine-grained elements. Finally, because this view represents only one of many possible application states, exploring the effectiveness of the interaction requires replicating the process for other views.

Unfortunately, exploring this same interaction by coding a low-fidelity interactive prototype capable of supporting it also entails considerable effort. In comparison to graphic design tools, low-fidelity interactive prototypes also make it considerably more difficult to examine alternative layouts, typefaces, controls, and other aspects of the design. Testing pixel-perfect versions of interactions and transitions in a coded prototype effectively requires committing to and implementing the entire design. 

Across the five projects, we often took both approaches, using graphic design tools to sketch and visually polish components and visualization views, while simultaneously implementing prototypes to test the impact of the interaction. We also used interactive charting tools like Tableau~\cite{Tableau} and RAWGraphs~\cite{mauri_rawgraphs:_2017} to create data-accurate visualization elements that could then be exported back into graphic and interaction design tools to create richer mockups. However, the gaps between each of these tools is considerable and interaction prototyping consumed a substantial amount of the design team's bandwidth.



\begin{figure}[tb]
  \centering
  \includegraphics[width=\columnwidth]{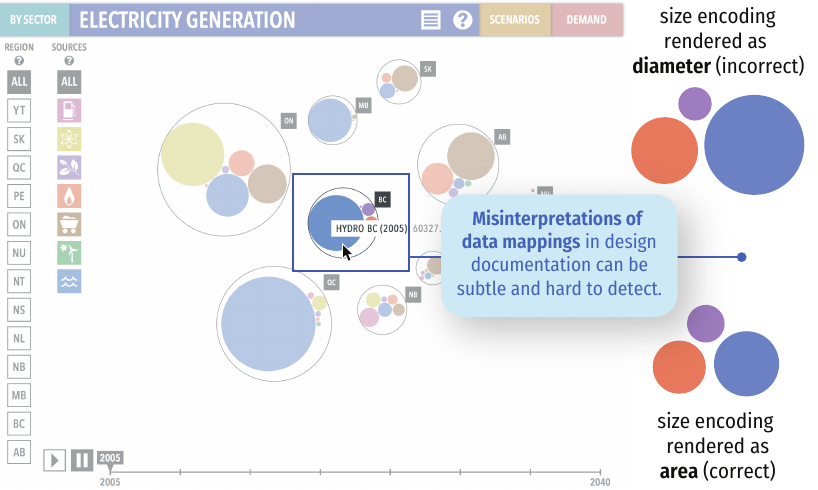}
  \caption{A static mockup of a bubble chart visualization. While tooltips on the bubbles provide precise values, the image alone does not contain enough information to enable a developer to re-create it. In particular, it does not specify that the data values shown are mapped to the area, not the diameter, of the circles. }
  \label{fig:energyfutures-bubblechart}
\end{figure}

\ChallengeSection{\ChallengeDataMapping}
\emph{Implementing a data mapping correctly requires more detail and precision than can be easily inferred from a mockup of a visualization view. However, precise and complete specification of data mappings is not well-supported by current design tools.}

The mapping from data to visual representation is the most fundamental aspect of a visualization. As such, it is critical that a designer creating a visual mapping be able to communicate their intent to others on the team, especially developers. Current options for communicating data mappings are limited and must be created manually. A static rendering of the view may seem sufficient if the visualization is designed to be easily read by non-experts. However\cam{,} example views may not capture many of the nuances and details that are important for implementation purposes. For example, correctly mapping a data value to a visual mark often requires data transformations or lookups, which may involve multiple hidden steps (such as using a classification algorithm to bin heat map values). Furthermore, the complexity of the data may make it difficult to explicitly show all data cases in the provided views, particularly where interaction is concerned (see Challenge~\CDataInteraction).

\begin{figure}[b]
  \centering
  \includegraphics[width=\columnwidth]{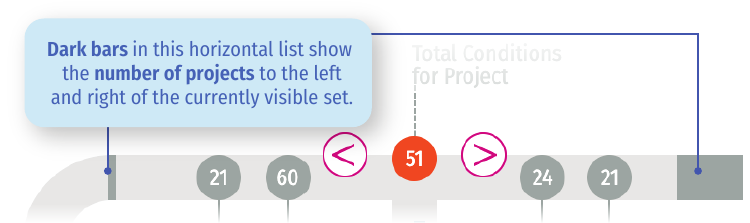}
  \caption{\add{The mapping between number of projects and the width of the bars in this horizontal list was unclear both in the design document and in follow-up conversations.}}
  \label{fig:datamapping-conditions}
\end{figure}

One clear example of this issue emerged late in the \textit{Energy Futures } project, which included the bubble chart visualization shown in Figure~\ref{fig:energyfutures-bubblechart}. The visualization design documentation outlining this visualization included one static view of the bubble chart created in Adobe Illustrator~\cite{illustrator}, together with information about which data columns were to be mapped to the size of the circles. The visualization designers expected, but did not precisely specify, that ``size'' would be interpreted as area rather than diameter. This expectation \cam{drew from} their own domain knowledge of common visualization guidelines. However, this guideline is not inherently obvious, particularly to non-experts. In this case, the initial implementation mapped the data to the circles' diameter rather than their area. This subtle difference in the mapping was difficult to detect via visual inspection alone (see Challenge~\CIterationsDiff). Ultimately, the error was caught by happenstance. However, once the discrepancy was noticed, extra development time was needed to change the mapping to what the designers originally intended. 

\add{Another example from the \textit{Pipeline Conditions} project illustrates that data mapping specification is challenging even in face-to-face conversation. One portion of \cam{this} visualization displayed a horizontal scrolling list of projects (Figure~\ref{fig:datamapping-conditions}). On either side of this list a bar denoted the number \cam{of} projects to the left and right. At one face-to-face design review session, the issue was raised of how best to map the number of remaining projects to the size of the bars, as the space allotted for the bars \cam{was} quite small. A simple solution was agreed upon verbally. However, the next implementation cycle revealed that, in fact, even this simple solution could be interpreted in multiple ways: one team understood that the maximum size of the bar would \cam{be set based on the entry with the largest number of possible items, while} the other team understood that the maximum size of the bar would represent a fixed value.} \cam{This confusion ultimately required a third set of follow-up discussions in order for the teams to reach a consensus.}


\ChallengeSection{\ChallengeIterationsDiff}

\emph{Because it is difficult to systematically compare implementations against design documents, there is a serious risk that misinterpretations or misapplications of the data mapping \cam{can} go unnoticed during and after development.}

As a design is implemented, differences can emerge due to a variety of factors including bugs, data updates, inconsistencies in the initial designs, and misinterpretations of the data mapping. This is in part because the implementation is separate from the design documentation. The primary method of testing the implementation's adherence to the design is visual inspection and comparison to the original specification. Furthermore, every new iteration of an implementation typically introduces numerous small changes. This can make it difficult to manually keep track of what has and has not been inspected or to know when it is the right time to flag an issue. Even if all parties recognize the importance of preserving the data mapping, visual inspection of the implementation alone may not reveal misinterpretations of the design or \cam{of} the data mapping at any given iteration.

\begin{figure}[tb]
  \centering
  \includegraphics[width=\columnwidth]{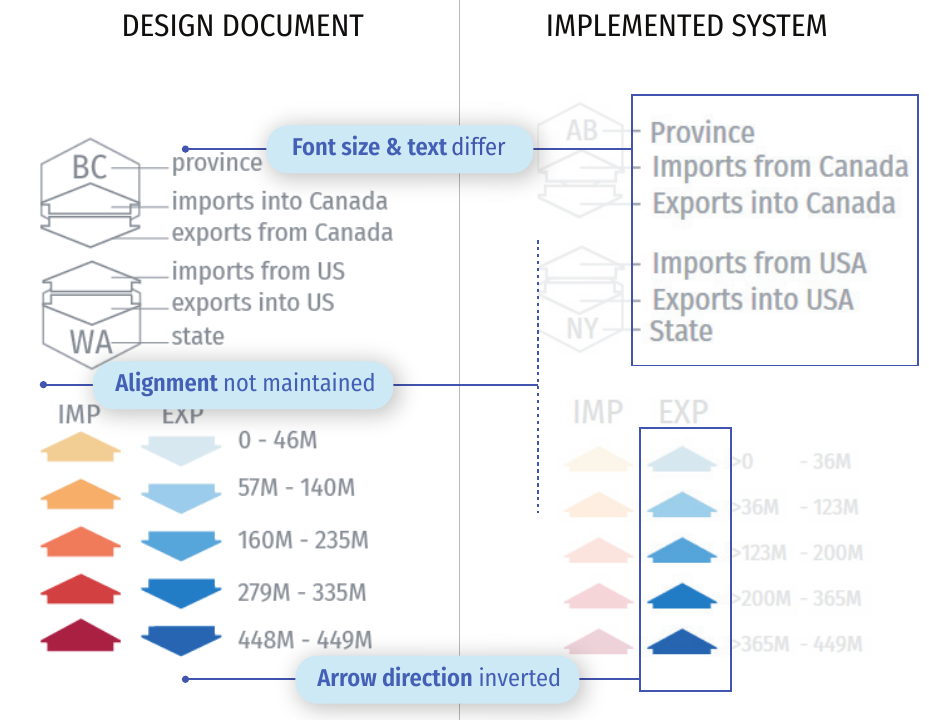}
  \caption{Detailed differences between a piece of a visualization design as specified in the original design documentation (left) and an implemented version of that design (right). Manually inspecting each iteration of the developed visualization to compare against the design document is tedious and error-prone.}
  \label{fig:diffs-legend}
\end{figure}

The example in Figure~\ref{fig:diffs-legend} shows a comparison between a small portion of a visualization design document and an implementation of that design. Many small differences related to both presentation design and the data mapping are apparent, including issues with the typography and alignment. In addition, the set of blue arrows in the legend point up, but should point down to align with the downward export arrows in the rest of the visualization.

The incorrectly-sized bubble charts described in Challenge~\CPDataMapping~also illustrate the difficulty of noticing mistakes in implemented designs. While detecting this \cam{kind of} error using only visual inspection is already difficult, it becomes even more challenging when there are frequent data updates during the visualization development process. Frequent changes and updates like these can lead to uncertainty about whether the visualization \cam{matches} expectations due to an incorrect mapping or simply due to differences in the data.


\section{Discussion}
\label{sec:discussion}

\add{The challenges we have highlighted stem directly from the intrinsic connection between visualizations and the source data that drives them. When compared \cam{to} other kinds of interfaces, data affects the visibility of elements, their layout, and their appearance to a much greater extent. Viewers of these tools interact not only with the interface but \emph{with the data}. As a result, developers and software engineers must also deal with the pragmatic limitations of the datasets when considering performance and interactive capabilities. Yet, for designers, anticipating all of the implications of scale and interaction for a given dataset remains challenging. This is \cam{magnified} by the reality that datasets are likely to be updated many times both during and after the design process. Together, these challenges suggest a variety of opportunities for research and tool creation that could specifically support the visualization design process\cam{,} both for individuals and \cam{for} collaborative teams.}

\subsection{Data Characterization}

The process of exploring and characterizing new datasets in preparation for visualization design work has much in common with data analysis, and therefore many existing data analysis tools can be appropriated within this space. However, Challenge~\CPDataUpdate~illustrates some opportunities to create data characterization tools and processes that are dedicated specifically to visualization design. Changes to data characteristics can have a substantial impact upon the final visualization outcome because visualization design choices typically reflect the shape and parameters of the given dataset. However, the extent to which new data may impact the robustness of the visualization may not be clear during the data update process. 

Data characterization tools could help remove this disconnect by helping designers better understand how the data has changed from one version to the next and how those changes might alter the design of the visualization. This could include tools for highlighting changes in data column names, extrema, and statistical distributions of data, or for simulating likely future changes based on the current distribution of values.
Recent work on semi-automated approaches for outlier detection and profiling in 
data mining toolboxes like Orange~\cite{demsar_orange_13} \cam{or} data discovery tools like DataTours~\cite{Mehta2017} may provide a useful starting point. Similarly, visual tools for quickly comparing the distribution of values in different datasets might help designers more readily detect problematic changes without relying on statistical summaries~\cite{Matejka_2017}.

\subsection{Design Phase}
\noindent \textbf{Data-Driven Visualization Ideation.} A number of the challenges we experienced are associated fundamentally with the challenge of ideating data-driven visualization designs. Existing commercial tools \add{for manual vector-based graphic design} such as Adobe Illustrator~\cite{illustrator} have little support for creating complex, data-driven views, while visualization \add{exploration and generation} tools such as Tableau~\cite{Tableau} or RAWGraphs~\cite{mauri_rawgraphs:_2017}
have limited support for custom visuals and interaction. Meanwhile, programming tools and lower-level libraries can be challenging to use as rapid ideation platforms and can disempower non-programmers. 
Fortunately, recent projects like Data Illustrator~\cite{liu_data_2018}, Data Ink~\cite{xia_dataink:_2018} and Data-Driven Guides~\cite{Kim_2017} highlight the potential for more expressive data-driven graphic design tools. 
Several of the challenges we experienced emphasize the need for further work in this space. More direct, dynamic, and expressive tools for designing with data could facilitate rapid exploration of different design alternatives even in the face of changing data~(\CDataUpdate). Similarly, more rapid exploration of data-driven design alternatives could make it easier to discover unexpected edge cases~(\CEdgeCases) and prototype data-dependent interactions (\CDataInteraction).
Mei et al.~\cite{Mei2017} identify several additional research directions for these kinds of tools, including supporting refinement based on existing visualizations, providing better debugging support, and exploring programming for dynamic data and interaction.


Where artifact creation cannot be unified into one single authoring platform, better handoff tools may also offer opportunities for designers and developers to synchronize the artifacts across multiple systems. 
Already, tools like Hanpuku~\cite{bigelow_iterating_2017} have explored bridging the graphic design expressivity of Adobe Illustrator and the data-driven prototyping capabilities of D3~\cite{Bostock2011}. However, designing bi-directional workflows between these sorts of existing tools usually entails compromise --- often intersecting the limitations of each tool and limiting the pieces of functionality that can be translated. This suggests that unidirectional handoff tools, like those now widely used in interaction design, are a more likely first step. 

\noindent \textbf{Data-Driven Interaction Prototyping.} Prototyping data-driven interactions within a visualization design is important for exploring different interaction options, ensuring the scalability and understandability of those interaction options, and communicating interaction designs to developers~(\CDataInteraction). Yet data-driven interactions can be complex to prototype because one interaction can simultaneously cause a change to a large number of data-driven elements in a design.
Unfortunately, commercially-available user interface prototyping tools do not address this challenge. \rem{While} \add{Commercial interaction design} tools \rem{like Adobe XD} make it straightforward to prototype interactions and transitions using  static mockups, \add{and even provide some limited support for data-driven prototyping --- for example by populating user profiles or lists of information. However,} they provide little support for creating \add{visualization views whose layout, appearance, and interactivity are all deeply and inherently driven by data.}  
As such, opportunities remain for new tools that allow designers to more expressively prototype and test potential interactions, either by bootstrapping on top of existing visualization tools or via new authoring interfaces.

\noindent \textbf{Data Mapping Documentation.} %
Communicating and documenting design intent is useful not only for explaining visualization designs to a development team, but also when communicating with other team members or stakeholders and when producing project documentation. One opportunity highlighted by challenge \CPDataMapping, is to design tools and methods for communicating data mappings. Such tools would support explicit communication of the relationship between data structures and their graphical representation with enough detail to convey any transformations, calculations, and algorithms required. 

Related work on visualization grammars ~\cite{Wilkinson_2010,Stolte,Satyanarayan_2017} provides a useful starting point, as do projects that represent the visualization pipeline~\cite{Tobiasz_2009} and support the deconstruction and modification of data mappings for existing visualizations~\cite{Harper2014}. However, there remains a need for data mapping tools that are accessible to designers without programming skills but which still communicate the nuances of a data mapping in enough detail to reproduce it programmatically.


\noindent \textbf{Data Visualization Design Documentation} %
Although data mappings are fundamental to a visualization, they are only one part of its design. Ultimately, any data mapping must be communicated as part of a larger body of design documentation that also captures graphical presentation (including layout, typography, and color) as well as interaction. While currently this documentation is often ad-hoc and informal, more systematic tools for capturing and communicating design details could be valuable in larger visualization design projects. Given the highly visual and interactive nature of visualization designs, one basis for these kinds of documents could be explorable explanations or other interactive documents. This format, popularized by Bret Victor~\cite{victor_humane_2014} is increasingly used in data analysis practice, and recent literate programming tools like Observable~\cite{observablehq}, litvis~\cite{wood_2019_litvis}, and Idyll~\cite{conlen_idyll_2018} may provide promising platforms for creating and documenting visualization designs.


Annotation tools can also play an instrumental role in design communication, particularly when aspects of a design may not be obvious from the  visual artifacts alone. Already, annotation is often present in authoring tools for low-fidelity user experience prototypes (cLuster \cite{perteneder_cluster:_2015}, D.Note \cite{hartmann_d.note:_2010}, SketchComm \cite{li_sketchcomm:_2012}, SILK \cite{landay_silk:_1996}). Going forward, visualization-specific annotation libraries like ChartAccent \cite{ren_chartaccent:_2017} have the potential to enable richer data-driven selection and annotation that could also support visualization design documents.

\subsection{Development Phase}
As emphasized by challenge \CPIterationsDiff, it is important that anyone testing an implemented visualization against its design documentation is able to identify discrepancies between the intended design and the implementation. Moreover, they should be able to differentiate between discrepancies that are due to data differences, discrepancies that are due to incomplete implementation, and unintentional discrepancies that are due to miscommunication or mistakes. The most vital discrepancies are those pertaining to data mappings (\CDataMapping). However, presentation discrepancies related to screen size, alignment, layout, typography, and non-data color issues like those illustrated in Figure~\ref{fig:diffs-legend} also play a role. In these cases, new tools for supporting differencing and visual comparison between visualizations~\cite{gleicher_visual_2011,gleicher_considerations_2018} may be particularly valuable. Interaction discrepancies, on the other hand, may be more difficult to detect, but are especially important when they can affect the understanding of the data. \add{Working visualization prototypes that simulate the final interactions can mitigate some of these issues, but only when the technical constraints of the design and development teams are aligned ahead of time (\CTechnicalChallenges}).


\subsection{Many Paths to Visualization Design}
The challenges for data visualization design and handoff described here are a direct reflection of our own experience in a large, multi-team environment \add{and are limited to our perspective as the design team}. With the exception of the first project, the physical and temporal separation between data compilation, visualization design, and development were relatively high. Not all visualization projects are configured in this way. Some projects involve multiple co-located people in highly specialized roles (for example in newsrooms~\cite{FischerBaum_2018}, where journalists, data analysts, designers and programmers might work closely together on very tight timelines). Meanwhile, many projects still involve individual designer-developers taking on many roles simultaneously.\rem{We see our contributions as adding value to this practitioner-oriented research space.}  \add{There is a wealth of literature discussing managing communication on design teams~\cite{eckert_communication_2005}, including work on developing shared mental models of tasks, teams, and processes~\cite{badke-schaub_mental_2007, johnson_measuring_2007}, sharing meaning-making activities~\cite{larsson_making_2003}, and managing the organization of design work~\cite{chiu_organizational_2002}. } As such, it is clear that technology is only one way to help mitigate communication challenges and that processes, communication frameworks, education, environment, mutual trust/comfort, and increased face-to-face time are all factors that affect the design process.

The separation of our teams likely \cam{amplified} and made it easier to identify some of the challenges we articulate. Our projects transitioned from phase to phase in very discrete stages and the need for clear communication and knowledge transfer between teams was very strong. However, \rem{we believe} the challenges we identify are more closely related to \add{the way that data permeates every aspect of the design process} \rem{the phases of the design process and the tools that support them, rather than the distribution of personnel}\add{and would remain, to some extent, regardless of team configuration}. \rem{While mechanisms for knowledge transfer between teams are useful, those same mechanisms could help even individual designer-developers as they transition from role to role in their projects. Addressing these challenges could be useful even for individual designer-developers. For example, a Consider, for example, the potential for a data mapping support tool could encourage a rigorous examination of the data and the data mapping and  be used as a basis for implementation and documentation. Similarly, a data characterization tool could formally support the exploratory process that any visualization designer performs when starting a design.}
As we have shown, data impacts nearly every facet of visualization design. A single update to the data distribution can completely change the effectiveness of \cam{a} visualization (\CDataUpdate). Incorporating standard data interactions into a design can increase the \cam{number of application states} that need to be \cam{considered} and make it much more difficult to find edge cases (\CEdgeCases). \add{Moreover, the fundamental dependence of the visualization on data may require a robust \cam{and complex backend implementation}, the constraints of which are not always apparent during the design phase, making it difficult to anticipate technical challenges or provide coded prototypes that can easily be translated to final implementations~(\CTechnicalChallenges).} Prototyping interactions with data can also involve tedious calculations and significant changes from view to view (\CDataInteraction).  Likewise, articulating and testing the mapping from data to its visual representation is a fundamental and complex task that is not well-supported by  conventional user interface design tools (\CDataMapping, \CIterationsDiff). \cam{Others before us have articulated difficulties in design handoff, including articulating edge cases~\cite{leiva2019enact}. However,} everything in the visualization design --- the mapping, the graphic design, and the interaction design --- depends \add{deeply} upon the data\cam{. T}herefore, the data adds complexity to every step of this process.s Finding and addressing opportunities to support that complexity could make the visualization design process more effective, more expressive, and more accessible to people with a variety of skillsets.

\section{Conclusion}
\add{Based on reflections on our experiences as designers during the data characterization, visualization design, and development phases of several large-scale collaborative visualization projects, we have highlighted the increased complexity that data fundamentally introduces to the design process. Data-related} challenges span all phases of design and include adapting to late-stage data changes, anticipating edge cases, articulating data-dependent interactions, communicating data mappings, and preserving data mapping integrity in implementation. These point to several opportunities to create tools that directly support the visualization design process through specific data-related features. Creating these more powerful tools could make the design process more robust, efficient, and accessible to people in a variety of design roles.

\acknowledgments{
We thank (alphabetically)\cam{:} Amanda Harwood, Annette Hester, Ryan Hum, Katherine Murphy, and Peter Watson and their teams at the National Energy Board of Canada (NEB) for facilitating and supporting this project. We also thank the software development team at VizworX. %
\cam{Finally, we} thank members of the design teams who took part in the projects 
\cam{that} led to this paper (alphabetically)\cam{:} %
Bon Adriel Aseniero, %
Peter Buk, %
Shreya Chopra, %
\cam{Claudio Esperan\c{c}a,} %
Tina Huynh, %
Lisa Hynes, %
Lindsay MacDonald Vermeulen, %
Claudia Maurer, %
Charles Perin, %
\cam{Richard} Pusch, %
Lien Quach, %
Katrina Tabuli, %
Markus Tessmann, %
Sarah Storteboom, %
Sonja Thoma, %
\cam{and} Jo Vermeulen.
}

\cam{This project was funded in part by The National Energy Board of Canada, the National Sciences and Engineering Research Council of Canada, Alberta Innovates Technology Futures, SMART Technologies, and the European Union’s Horizon 2020 research and innovation programme under the Marie Sklodowska-Curie grant agreement No. 753816.}

\bibliographystyle{abbrv-doi}
\bibliography{bibliography/converted_to_latex.bib}


\end{document}